\documentstyle[amsmath,amssymb,epsfig,12pt]{article} 
\begin{document} 
\setlength{\parskip}{0.45cm} 
\setlength{\baselineskip}{0.75cm} 
%
%
%
\begin{titlepage} 
\setlength{\parskip}{0.25cm} 
\setlength{\baselineskip}{0.25cm} 
\begin{flushright} 
DO-TH 2001/01\\ 
\vspace{0.2cm} 
January 2001 
\end{flushright} 
\vspace{1.0cm} 
\begin{center} 
\LARGE 
{\bf Spin-Dependent Structure Functions}\\ 
\LARGE{\bf of the Photon} 
\vspace{1.5cm} 
 
\large 
M. Gl\"uck, E.\ Reya, C.\ Sieg\\ 
\vspace{1.0cm} 
 
\normalsize 
{\it Universit\"{a}t Dortmund, Institut f\"{u}r Physik,}\\ 
{\it D-44221 Dortmund, Germany} \\ 
\vspace{0.5cm}

\vspace{1.5cm} 
\end{center} 
 
\begin{abstract} 
\noindent 
The implications of the positivity constraint on the presently unknown 
polarized structure function of the photon, $g_1^{\gamma}(x,Q^2)$, are 
studied in detail.  In particular the non--trivial consequences of this 
constraint, $|g_1^{\gamma}(x,Q^2)|\leq F_1^{\gamma}(x,Q^2)$, in the 
next--to--leading order analysis of $g_1^{\gamma}$ are pointed out by 
employing appropriate (DIS) factorization schemes related to  
$g_1^{\gamma}$ 
and $F_1^{\gamma}$ (rather than to $F_2^{\gamma}$). 
\end{abstract} 
\end{titlepage} 
 
 
 
The spin dependent structure function $g_1^{\gamma}(x,Q^2)$ of a  
longitudinally polarized photon was studied \cite{ref1,ref2} within 
the framework of the radiative parton model, developed \cite{ref3} for 
the presently well measured and known structure function $F_2^{\gamma} 
(x,Q^2)$ of an unpolarized photon.  In particular the next--to--leading 
order (NLO) analysis \cite{ref2} of $g_1^{\gamma}$ adopted a perturbatively 
stable DIS$_{\gamma}$ factorization scheme, as advocated in \cite{ref3}, 
and implemented some boundary conditions \cite{ref1} at the low input 
scale $Q^2=\mu^2\simeq 0.3$ GeV$^2$ of the radiative parton model. 
These boundary conditions led, however, to a {\underline{violation}} 
of the positivity constraint 
\begin{equation} 
|A_1^{\gamma}(x,Q^2)|\equiv|g_1^{\gamma}(x,Q^2)/F_1^{\gamma}(x,Q^2)|\leq 1\, . 
\end{equation} 
It was therefore suggested \cite{ref2} to repeat the analysis \cite{ref3} 
of $F_2^{\gamma}$ in a DIS$_{\gamma}$ factorization scheme related to 
$F_1^{\gamma}$ rather than to $F_2^{\gamma}$ which was the source of 
the above mentioned violation.  Such a reanalysis is obviously rather 
time consuming and leads, moreover, to a diminished perturbative stability 
of the resulting parton distributions.  In the present letter we propose 
an alternative solution to the positivity constraint which avoids the 
need for the above mentioned reanalysis of the data on $F_2^{\gamma}$. 
 
For this purpose we recall that the DIS$_{\gamma}$ factorization scheme, 
suggested and adopted in \cite{ref3} for unpolarized photon structure 
functions, is related to $F_2^{\gamma}$ as given in  
NLO(${\overline{\rm MS}}$) by  
\begin{eqnarray} 
\frac{1}{x}\,F_2^{\gamma}(x,Q^2) & = & \sum_{q=u,d,s} e_q^2 
   \biggl\{ q^{\gamma}(x,Q^2)+\bar{q}\,^{\gamma}(x,Q^2) \nonumber\\ 
& & + \frac{\alpha_s(Q^2)}{2\pi}  
   \left[C_{q,2}\otimes(q+\bar{q})\,^{\gamma}  
          + 2 C_{g,2}\otimes g^{\gamma} \right] 
    + 2e_q^2\,\frac{\alpha}{2\pi}\, C_{\gamma,2}(x) \biggr\} 
\end{eqnarray} 
\vspace{-0.6cm} 
 
\noindent where $\otimes$ denotes the usual convolution integral, and  
$\bar{q}\,^{\gamma}(x,Q^2) = q^{\gamma}(x,Q^2)$ and $g^{\gamma}(x,Q^2)$ 
provide the so--called  `resolved' contributions of $\gamma$ to 
$F_2^{\gamma}$ with the usual ${\overline{\rm MS}}$ coefficient  
functions 
\begin{eqnarray} 
C_{q,2}(x) & = & C_{q,1}(x)+\frac{4}{3}\,2x\nonumber\\ 
& = & \frac{4}{3}\left[ (1+x^2)\left( \frac{\ln(1-x)}{1-x}\right)_+ 
    -\frac{3}{2}\ \frac{1}{(1-x)_+} - \frac{1+x^2}{1-x}\, \ln x\right. 
     \nonumber\\ 
& &  \left. +3+2x 
     -\left( \frac{9}{2}+\frac{\pi^2}{3}\right) \delta(1-x)\right] 
      \nonumber\\ 
C_{g,2}(x) & = & C_{g,1}(x)+\frac{1}{2}\,4x(1-x)\nonumber\\ 
 & = & \frac{1}{2}\left\{ \left[ x^2+(1-x)^2\right]\, \ln\,  
       \frac{1-x}{x} +8x(1-x)-1\right\}\,, 
\end{eqnarray} 

while $C_{\gamma,2}$ provides the  `direct' contribution as calculated 
according to the `box' diagram $\gamma^*(Q^2)\gamma\to q\bar{q}$ : 
\begin{equation} 
C_{\gamma,\,i}(x)=\frac{3}{(1/2)}\,C_{g,\,i}(x) 
\end{equation} 
with $i=1,2$.  We have suppressed in (2) the contributions from heavy 
($c,\, b$) quarks since they are irrelevant for our present considerations. 
(The $C_1$ coefficient functions refer to $F_1^{\gamma}$ needed below.) 
Notice that in {\underline{un}}polarized photon (and proton) DIS it is 
common to use the `mixed' structure function $F_2=2xF_1+F_L$, rather 
than the purely transverse $F_1$ structure function, since the measured 
cross section is, apart from kinematically suppressed contributions,  
directly proportional to $F_2$.  In order to avoid the instabilities 
encountered in NLO($\overline{\rm MS}$) in the large--$x$ region due to 
the $\ln(1-x)$ term in $C_{\gamma}$ in (4), the entire `direct'  
$C_{\gamma,2}$ term in (2) is absorbed into the $\overline{\rm MS}$ 
(anti)quark densities $q^{\gamma}=\bar{q}\,^{\gamma}$ in (2) which 
defines the so--called DIS$_{\gamma}$ factorization scheme \cite{ref3}: 
Eq.(5)
\begin{eqnarray} 
q^{\gamma}(x,Q^2)_{\rm DIS_{\gamma}} & = & q^{\gamma}(x,Q^2) +  
     e_q^2\, \frac{\alpha}{2\pi}\, C_{\gamma,2}(x)\nonumber\\ 
g^{\gamma}(x,Q^2)_{\rm DIS_{\gamma}} & = & g^{\gamma}(x,Q^2)\, . 
\end{eqnarray} 
This redefinition of parton distributions implies that the  
NLO($\overline{\rm MS}$) splitting functions $k_{q,g}^{(1)}(x)$  
of the photon into quarks and gluons, appearing in the inhomogeneous  
NLO renormalization group (RG)$Q^2$--evolution equations \cite{ref3}  
for $f^{\gamma}(x,Q^2)$, have correspondingly to be transformed  
according to \cite{ref3,ref4} 
\begin{eqnarray} 
k_q^{(1)}(x)_{\rm DIS_{\gamma}} & = & k_q^{(1)}(x) - e_q^2\, P_{qq}^{(0)} 
   \otimes C_{\gamma,2}\nonumber\\ 
k_g^{(1)}(x)_{\rm DIS_{\gamma}} & = & k_g^{(1)}(x) - 2 \sum_q 
    e_q^2\, P_{gq}^{(0)}\otimes C_{\gamma,2}  
\end{eqnarray} 
\vspace{-0.5cm} 
where 
\begin{eqnarray} 
k_q^{(1)}(x) & = &\frac{1}{2}\, 3 e_q^2\, \frac{4}{3} 
   \biggl\{ -(1-2x)\,\ln^2x -(1-4x)\,\ln\,x + 4\,\ln(1-x)-9x+4 \nonumber\\ 
& & +  \left[ x^2+(1-x)^2\right] \, \biggl[2\, \ln^2 x+ 2\, \ln^2(1-x)+ 
       4\, \ln\, x -4\, \ln\, x\, \ln(1-x)\nonumber\\ 
& &  -4\, \ln(1-x) + 10 
          -\frac{2}{3}\,\pi^2\biggr] \biggr\}\nonumber\\ 
k_g^{(1)}(x) & = & 3\sum_q e_q^2\frac{4}{3} 
    \left\{ -2(1+x)\ \ln^2 x-(6+10x)\, \ln\, x +\frac{4}{3x} + 
       \frac{20}{3}x^2+8x -16\right\} 
\end{eqnarray} 
with $k_q^{(1)}$ referring to each single (anti)quark flavor.  The 
LO splitting functions are given by $P_{qq}^{(0)}=\frac{4}{3}\left( 
\frac{1+ x^2}{1-x}\right)_+$ and $P_{gq}^{(0)}=\frac{4}{3}\left[ 
1+(1-x)^2\right]/x$. The NLO expression for $F_2^{\gamma}$ in the  
DIS$_{\gamma}$ factorization scheme is thus given by (2) with  
$C_{\gamma,2}$ being {\underline{dropped}}. 
 
In order to comply with the positivity constraint (1) for the polarized 
structure function $g_1^{\gamma}$ one has to consider a corresponding 
factorization scheme, DIS$_{\gamma,1}$, related to $F_1^{\gamma}$, the 
spin--averaged analogon to $g_1^{\gamma}$, which is given in  
NLO($\overline{\rm MS}$) by  
\begin{eqnarray} 
F_1^{\gamma}(x,Q^2) & = & \frac{1}{2}\sum_{q=u,d,s} e_q^2  
    \biggl\{ q^{\gamma}(x,Q^2)+\bar{q}\,^{\gamma}(x,Q^2)\nonumber\\ 
& & +\frac{\alpha_s(Q^2)}{2\pi}\, 
    \left[C_{q,1}\otimes(q+\bar{q})^{\gamma}+2C_{g,1}\otimes 
         g^{\gamma}\right] 
    + 2e_q^2\, \frac{\alpha}{2\pi}C_{\gamma,1}(x)\biggr\} 
\end{eqnarray} 
with the $C_1$ coefficient functions being given by eqs.\ (3) and (4). 
Absorbing now the entire  `direct' $C_{\gamma,1}$ term into the  
$\overline{\rm MS}$ quark densities $q^{\gamma}=\bar{q}\,^{\gamma}$ 
defines the DIS$_{\gamma,1}$ factorization scheme: 
\vspace{-0.5cm} 
\begin{eqnarray} 
q^{\gamma}(x,Q^2)_{\rm{DIS}_{\gamma,1}} & = & q^{\gamma}(x,Q^2)+ 
      e_q^2\, \frac{\alpha}{2\pi}\, C_{\gamma,1}(x)\nonumber\\ 
g^{\gamma}(x,Q^2)_{\rm{DIS}_{\gamma,1}} & = & g^{\gamma}(x,Q^2) 
\end{eqnarray} 
with the corresponding change of the NLO($\overline{\rm MS}$) photon 
splitting functions $k_{q,g}^{(1)}(x)$, appearing in the  
NLO($\overline{\rm MS}$) RG evolution equations, 
\begin{eqnarray} 
k_q^{(1)}(x)_{\rm{DIS}_{\gamma,1}} & = & k_q^{(1)}(x) -e_q^2\,  
       P_{qq}^{(0)}\otimes C_{\gamma,1}\nonumber\\ 
k_g^{(1)}(x)_{\rm{DIS}_{\gamma,1}} & = & k_g^{(1)}(x)-2\sum_q e_q^2\, 
       P_{gq}^{(0)}\otimes C_{\gamma,1} 
\end{eqnarray} 
in contrast to eq.\ (6). From the definitions (5) and (9) one obtains: 
\begin{eqnarray} 
q^{\gamma}(x,Q^2)_{\rm{DIS}_{\gamma,1}} & = & q^{\gamma} 
   (x,Q^2)_{\rm{DIS}_{\gamma}} + e_q^2\, \frac{\alpha}{2\pi} 
    \left[C_{\gamma,1}(x)-C_{\gamma,2}(x)\right]\nonumber\\ 
& = & q^{\gamma}(x,Q^2)_{\rm{DIS}_{\gamma}} - e_q^2\,  
      \frac{\alpha}{2\pi}\, 12x(1-x)\nonumber\\ 
g^{\gamma}(x,Q^2)_{\rm{DIS}_{\gamma,1}} & = & g^{\gamma} 
                 (x,Q^2)_{\rm{DIS}_{\gamma}} 
\end{eqnarray} 
Thus the NLO expression for $F_1^{\gamma}$ in the DIS$_{\gamma,1}$ 
factorization scheme is given by (8) with the $C_{\gamma,1}$ term 
being {\underline{dropped}}.  Furthermore, the parton distributions 
in the DIS$_{\gamma,1}$ scheme are uniquely determined in terms of 
the well known DIS$_{\gamma}$ distributions \cite{ref5,ref6} in  
eq.\ (11), $f^{\gamma}(x,Q^2)_{\rm{DIS}_{\gamma}}$ with $f=u,\,d,\,s,\,g$, 
related to $F_2^{\gamma}$.  Since the perturbative stability has been 
optimized \cite{ref5,ref6} with respect to the experimentally 
measured structure function $F_2^{\gamma}$, the stability may  
obviously be diminished at some other place, e.g.\ $F_1^{\gamma}$, 
where the difference between the leading order (LO) predictions and 
the NLO ones is somewhat more pronounced as compared to  
$F_2^{\gamma}$.  This of course does {\underline{not}} affect the 
reliability of the {\underline{NLO}} predictions.  (The LO expressions 
for $F_{1,2}^{\gamma}$ are obviously obtained from eqs.\ (2) and (8) 
by simply setting $C_{q,g,\gamma}=0$.) 
 
The polarized parton distributions $\Delta f^{\gamma}$ in the analogous 
DIS$_{\Delta\gamma}$ factorization scheme are obtained in a similar 
way by considering the spin--dependent structure function  
$g_1^{\gamma}$ in (1) which, for the light $u,\,d,\,s$ quarks, is 
in NLO($\overline{\rm MS}$) given by 
\begin{eqnarray} 
g_1^{\gamma}(x,Q^2) & = & \frac{1}{2}\sum_{q=u,d,s} e_q^2 
  \biggl\{ \Delta q^{\gamma}(x,Q^2) 
     +\Delta\bar{q}\,^{\gamma}(x,Q^2)\nonumber\\ 
& & + \frac{\alpha_s(Q^2)}{2\pi}  
   \left[ \Delta C_q\otimes \Delta(q+\bar{q})^{\gamma} + 
     2\Delta C_g\otimes\Delta g^{\gamma}\right] 
      + 2 e_q^2\, \frac{\alpha}{2\pi}\, \Delta C_{\gamma}(x)\biggr\} 
\end{eqnarray} 
with $\Delta\bar{q}\,^{\gamma}=\Delta q^{\gamma}=q_+^{\gamma} 
-q_-^{\gamma}$ and $\Delta g^{\gamma}=g_+^{\gamma}-g_-^{\gamma}$ 
as compared to the spin--averaged $\bar{q}\,^{\gamma}=q^{\gamma} = 
q_+^{\gamma}+ q_-^{\gamma}$ and $g^{\gamma}=g_+^{\gamma}+ 
g_-^{\gamma}$ in $F_1^{\gamma}$ in (8) in terms of the positive 
and negative helicity densities $q_{\pm}^{\gamma}$ and 
$g_{\pm}^{\gamma}$.  The polarized NLO($\overline{\rm MS}$) 
partonic coefficient functions \cite{ref7,ref8} for the `resolved' 
contributions of a longitudinally polarized photon are given by 
\vspace{0.3cm} 
\begin{eqnarray} 
\Delta C_q(x) & = & \frac{4}{3} \left[ (1+x^2) \left(  
 \frac{\ln (1-x)}{1-x}\right)_+ -\frac{3}{2}\, \frac{1}{(1-x)}_+ 
  -\frac{1+x^2}{1-x}\ln x\right.\nonumber\\ 
& & \left. +2+x -\left(\frac{9}{2}+\frac{\pi^2}{3}\right) 
   \delta(1-x)\right]\nonumber\\ 
\Delta C_g(x) & = & \frac{1}{2}\left[ (2x-1) \left( \ln 
   \frac{1-x}{x}-1\right) + 2(1-x)\right]\, , 
\end{eqnarray} 
\vspace{-0.8cm} 
 
\noindent and the  `direct' contribution of the polarized photon follows from 
\begin{equation} 
\Delta C_{\gamma}(x) = \frac{3}{(1/2)} \, \Delta C_g(x)\,. 
\end{equation} 
Absorbing this latter contribution in (12) entirely into the 
polarized (anti)quark distributions, one obtains, in complete 
analogy to the DIS$_{\gamma,1}$ scheme in (9), the polarized  
DIS$_{\Delta\gamma}$ factorization scheme \cite{ref2}, 
\begin{eqnarray} 
\Delta q^{\gamma}(x,Q^2)_{\rm{DIS}_{\Delta\gamma}} & =  
  & \Delta q^{\gamma}(x,Q^2)+e_q^2\, \frac{\alpha}{2\pi} 
      \Delta C_{\gamma}(x)\nonumber\\ 
\Delta g^{\gamma}(x,Q^2)_{\rm{DIS}_{\Delta\gamma}} & = 
  & \Delta g^{\gamma}(x,Q^2)\, . 
\end{eqnarray} 
Correspondingly, the polarized NLO($\overline{\rm MS}$) splitting 
functions $\Delta k_{q,g}^{(1)}(x)$ of the polarized photon into quarks 
and gluons, appearing in the inhomogeneous NLO RG $Q^2$--evolution 
equations \cite{ref2}, have to be changed according to 
\begin{eqnarray} 
\Delta k_q^{(1)}(x)_{\rm{DIS}_{\Delta\gamma}} & = &  
  \Delta k_q^{(1)}(x)-e_q^2\, \Delta P_{qq}^{(0)} 
    \otimes \Delta C_{\gamma}\nonumber\\ 
\Delta k_g^{(1)}(x)_{\rm{DIS}_{\Delta\gamma}} & = & 
   \Delta k_g^{(1)}(x) - 2 \sum_q e_q^2\, \Delta P_{gq}^{(0)} 
    \otimes \Delta C_{\gamma} 
\end{eqnarray} 
where \cite{ref2} 
\begin{eqnarray} 
\Delta k_q^{(1)}(x) & = & \frac{1}{2}\,3 e_q^2 \,\frac{4}{3} 
   \biggl\{ -9\ln x +8(1-x)\, \ln(1-x)+27x -22\nonumber\\ 
& & +(2x-1)\left[\ln^2 x+2\,\ln^2(1-x)-4\,\ln x\,\ln(1-x) 
    -\frac{2}{3} \pi^2\right]\biggr\}\nonumber\\ 
\Delta  k_g^{(1)}(x) & = & 3\sum_q e_q^2\, \frac{4}{3} 
   \left\{-2(1+x)\,\ln^2 x +2(x-5)\, \ln x -10(1-x)\right\} 
\end{eqnarray} 
with $\Delta k_q^{(1)}$ referring again to each single (anti)quark 
flavor and $\Delta P_{qq}^{(0)}=P_{qq}^{(0)}$, $\Delta P_{gq}^{(0)} 
=\frac{4}{3}(2-x)$.  The NLO expansion for $g_1^{\gamma}$ in the  
DIS$_{\Delta\gamma}$ scheme is thus given by (12) with the  
$\Delta C_{\gamma}$ term being {\underline{dropped}}. 
 
Following refs.\ \cite{ref1,ref2}, we shall now study two extreme 
scenarios : 
 
\noindent (i)$\quad$ a `maximal' scenario corresponding to an 
input 
\begin{equation} 
\Delta f^{\gamma}(x,\mu^2)_{\rm{DIS}_{\Delta\gamma}} = 
   f^{\gamma}(x,\mu^2)_{\rm{DIS}_{\gamma,1}}\, ; 
\end{equation} 
\noindent (ii)$\quad$ a  `minimal' scenario corresponding to an 
input 
\begin{eqnarray} 
\Delta q^{\gamma}(x,\mu^2)_{\rm{DIS}_{\Delta\gamma}} & = & 
  e_q^2\, \frac{\alpha}{2\pi} 
   \left[ C_{\gamma,1}(x)-C_{\gamma,2}(x)\right]\nonumber\\ 
& = & -e_q^2\, \frac{\alpha}{2\pi}\, 12x(1-x)\nonumber\\ 
\Delta g^{\gamma}(x,\mu^2)_{\rm{DIS}_{\Delta\gamma}} & = & 0 
\end{eqnarray} 
which derives from (11) for the minimal (`pointlike') boundary 
condition $f^{\gamma}(x,\mu^2)_{\rm{DIS}_{\gamma}}=0$ of the  
unpolarized photonic parton distributions in the DIS$_{\gamma}$ 
scheme \cite{ref3,ref6}.  Notice that (19) differs from the  
minimal (`pointlike') input  
$\Delta f^{\gamma}(x,\mu^2)_{\rm{DIS}_{\Delta\gamma}}=0$ 
considered in \cite{ref2}.  In order to facilitate a direct 
comparison with the results obtained in \cite{ref2} we shall also  
use the older GRV$_{\gamma}$ results \cite{ref5} for the 
unpolarized $f^{\gamma}(x,\mu^2)_{\rm{DIS}_{\gamma}}$ distributions 
in the DIS$_{\gamma}$ factorization scheme, which refer to a NLO 
input scale $\mu^2=0.3$ GeV$^2$, and which uniquely fix 
$f^{\gamma}(x,\mu^2)_{\rm{DIS}_{\gamma,1}}$ in (18) via eq.\ (11). 
(Our main conclusions remain unchanged, if we use the more recent 
unpolarized photonic parton distributions of \cite{ref6}.)  
In LO the `maximal' input (18) refers just to the common 
(scheme--independent) LO distributions \cite{ref5}, whereas the 
`minimal' input obviously implies, instead of (19), also a  
vanishing quark--input, i.e.\ $\Delta f^{\gamma}(x,\mu^2)_{\rm LO}=0$ 
which coincides with the `pointlike' solution for  
$\Delta f^{\gamma}(x,Q^2)$ and with the input of \cite{ref2} where 
\cite{ref5} $\mu_{\rm LO}^2 = 0.25$ GeV$^2$.   
   
In fig.\ 1 we show our maximal and minimal NLO results for  
$g_1^{\gamma}$ at a typical scale of $Q^2=10$ GeV$^2$ as obtained 
from the maximal and minimal input scenarios in (18) and (19), 
respectively, which fall somewhat below the results of \cite{ref2} 
as expected.  A comparison with $F_1^{\gamma}$ shows furthermore 
that the fundamental positivity constraint (1), $|g_1^{\gamma}| 
\leq F_1^{\gamma}$, is fulfilled throughout the {\underline{entire}} 
$x$--region (at any $Q^2$), in contrast to the violation of (1) 
observed in \cite{ref2}. The corresponding LO and NLO results for the 
asymmetry $A_1^{\gamma}=g_1^{\gamma}/F_1^{\gamma}$ is shown in fig.\ 2 
for two representative values of $Q^2$.  It should be noticed that  
at very large values of $x$  
($x$ \raisebox{-0.1cm}{$\stackrel{>}{\sim}$} 0.9) 
the numerical NLO results in figs.\ 1 and 2 become unreliable due to 
the influence of sizeable spurious ${\cal{O}}(\alpha_s,\, \alpha_s^2)$ 
terms \cite{ref3,ref5} encountered in the convolutions appearing in  
eqs.\ (8) and (12).  The `maximal' and `minimal' photonic parton 
distributions $\Delta u^{\gamma}$ and $\Delta g^{\gamma}$ are  
displayed in fig.\ 3 in LO and NLO(DIS$_{\gamma,1}$) at $Q^2=10$ 
GeV$^2$ and the corresponding asymmetries $A_f^{\gamma}\equiv 
\Delta f^{\gamma}/f^{\gamma}$ are shown in fig.\ 4, where we have 
again used the unpolarized $f^{\gamma}$ distributions from \cite{ref5} 
in order to facilitate a comparison with \cite{ref2}.  Our results 
for $\Delta u^{\gamma}$ and $\Delta g^{\gamma}$ in fig.\ 3 are  
similar to the ones in \cite{ref2}, with a larger difference between 
our LO and NLO predictions according to our different NLO inputs 
(18) and (19) which refer to the unpolarized DIS$_{\gamma,1}$ 
factorization scheme.  This, however, is irrelevant as discussed 
above for the reliability of the NLO predictions for the  
experimentally directly observable structure functions $g_1^{\gamma}$ 
and $F_1^{\gamma}$. 
 
In LO QCD, where cross sections (structure functions) are directly 
related to parton densities, the positivity constraint (1) for 
structure functions implies 
\begin{equation} 
|\Delta f^{\gamma}(x,Q^2)|\leq f^{\gamma}(x,Q^2) 
\end{equation} 
which is satisfied, $|A_{u,g}^{\gamma}|\leq 1$, as shown in fig.\ 4 
by the dashed curves.  At NLO, however, a simple relation between 
parton densities and cross sections no longer holds.  Parton 
distributions are renormalization and factorization scheme dependent 
objects; although universal, they are not physical, i.e.\ not 
directly observable.  Hence there are NLO contributions  
which may violate (20) in specific cases \cite{ref9}.  Such a  
curiosity occurs for our photonic parton densities which, for medium 
to large values of $x$, are dominated by the photon's splitting 
functions $(\Delta)k_{q,g}$ appearing as inhomogeneous terms 
in the RG evolution equations \cite{ref2,ref3,ref4}.  Up to NLO 
they are given by 
\vspace{0.3cm} 
\begin{equation} 
(\Delta)k_i(x,Q^2)=\frac{\alpha}{2\pi}(\Delta)k_i^{(0)}(x) +   
   \frac{\alpha\, \alpha_s(Q^2)}{(2\pi)^2}\, (\Delta)k_i^{(1)}(x) 
\end{equation} 
\vspace{-0.8cm} 
 
\noindent where in LO $(\Delta)k_q^{(0)}=\frac{1}{2}\, 3e_q^2 2\left[x^2 
\stackrel{+}{_{(-)}} (1-x)^2\right]$, while the NLO two--loops unpolarized 
splitting functions are given by (6), (7) or (10), and their polarized 
counterparts by (16) or (17), depending on the choice of the factorization 
scheme.  Our NLO results for $\Delta u^{\gamma}$ and $\Delta d^{\gamma}$ 
still satisfy the positivity constraint (20) as demonstrated by the  
solid curves for $A_u^{\gamma}$ in fig.\ 4 since in LO  
$|\Delta k_q^{(0)}|\leq k_q^{(0)}$ despite the fact that the  
subleading NLO contributions in general {\underline{violate}}  
$|\Delta k_q^{(1)}/k_q^{(1)}|\leq 1$.  The NLO gluon distributions, 
however, {\underline{violate}} (20) since the LO terms in (21) 
obviously vanish \cite{ref2,ref3}, $k_g^{(0)} = \Delta k_g^{(0)}=0$, 
and the dominant NLO terms $(\Delta)k_g^{(1)}$ in (21)  
{\underline{violate}} $|\Delta k_g^{(1)}/k_g^{(1)}|\leq 1$.  This 
violation of the NLO gluon  `positivity' is illustrated by the solid 
curves in fig.\ 4 for $A_g^{\gamma}$ where $A_g^{\gamma}>1$ for  
$x$ \raisebox{-0.1cm}{$\stackrel{>}{\sim}$} 0.6 
and 0.9 for the maximal and minimal scenario, respectively. 
 
To summarize, our approach to the positivity constraint (1) on the  
polarized structure function of the photon, $g_1^{\gamma}$, in NLO 
was to consider appropriate factorization schemes DIS$_{\gamma,1}$ 
and DIS$_{\Delta\gamma}$ which are naturally associated with the 
structure functions $F_1^{\gamma}$ and $g_1^{\gamma}$, respectively. 
Utilizing these factorization schemes we have been able to use the 
well established \cite{ref5,ref6} unpolarized NLO parton distributions 
of the photon, as given in the DIS$_{\gamma}$ factorization scheme 
associated with $F_2^{\gamma}$, in two different hypothetical 
`maximal' and `minimal' scenarios for the presently unknown  
$g_1^{\gamma}(x,Q^2)$.  We have thus shown that the time consuming 
NLO reanalysis of the data on $F_2^{\gamma}$ in the DIS$_{\gamma,1}$ 
factorization scheme, as proposed in \cite{ref2}, can in fact be  
avoided.  It turns out that our positivity respecting hypothetical  
NLO scenarios differ from their corresponding counterparts in  
\cite{ref2} thus illustrating the importance of a consistent  
implementation of the non--trivial NLO positivity constraint on  
$g_1^{\gamma}$. 
 
%
 
\newpage

\noindent{\large{\bf{\underline{Figure Captions}}}} 
\begin{itemize} 
\item[\bf{Fig.\ 1}.]  NLO predictions of the polarized photon structure 
      function $g_1^{\gamma}$ for the `maximal' and `minimal' inputs in 
      (18) and (19) with the unpolarized photonic parton distributions 
      $f_{\rm{DIS}_{\gamma,1}}^{\gamma}$ being calculated according to 
      the NLO(DIS$_{\gamma}$) distributions of ref.\ \cite{ref5} (in 
      order to facilitate a direct comparison with the results of ref.\ 
      \cite{ref2})$\,$.  These latter distributions determine also the  
      unpolarized photon structure function $F_1^{\gamma}$ in (8). 
 
\item[\bf{Fig.\ 2}.]  The spin asymmetry $A_1^{\gamma}\equiv g_1^{\gamma} 
      /F_1^{\gamma}$ in LO and NLO for the `maximal' and `minimal'  
      scenarios using the input distributions as in fig.\ 1 at two 
      representative values of $Q^2$. 
 
\item[\bf{Fig.\ 3}.]  Predictions for the NLO polarized photonic parton 
      densities in the DIS$_{\Delta\gamma}$ scheme, using the `maximal' 
      and `minimal' inputs in (18) and (19) referring to the unpolarized 
      DIS$_{\gamma,1}$ scheme where we use again the NLO(DIS$_{\gamma}$) 
      distributions of ref.\ \cite{ref5} for calculating the  
      $f_{\rm{DIS}_{\gamma,1}}^{\gamma}$ input densities in (18), in  
      order to facilitate a direct comparison with the results of ref.\ 
      \cite{ref2}.  For comparison we also show the corresponding LO 
      results with the input $f^{\gamma}(x,\mu^2)_{\rm LO}$ in (18) being 
      taken from ref.\ \cite{ref5} for the  `maximal' scenario, and where 
      `minimal' scenario obviously implies, instead of (19), also a 
      vanishing quark input, i.e.\ $\Delta f^{\gamma}(x,\mu^2)_{\rm LO}=0$. 
 
\item[\bf{Fig.\ 4}.] The parton spin--asymmetries $A_f^{\gamma}\equiv 
      \Delta f^{\gamma}/f^{\gamma}$ in LO and NLO  at $Q^2=10$ GeV$^2$ 
      for the `maximal' and 
      `minimal' scenarios using the input distributions as in fig.\ 3. 
\end{itemize} 


\pagestyle{empty}

\begin{figure}
\begin{center}
\epsfig{file=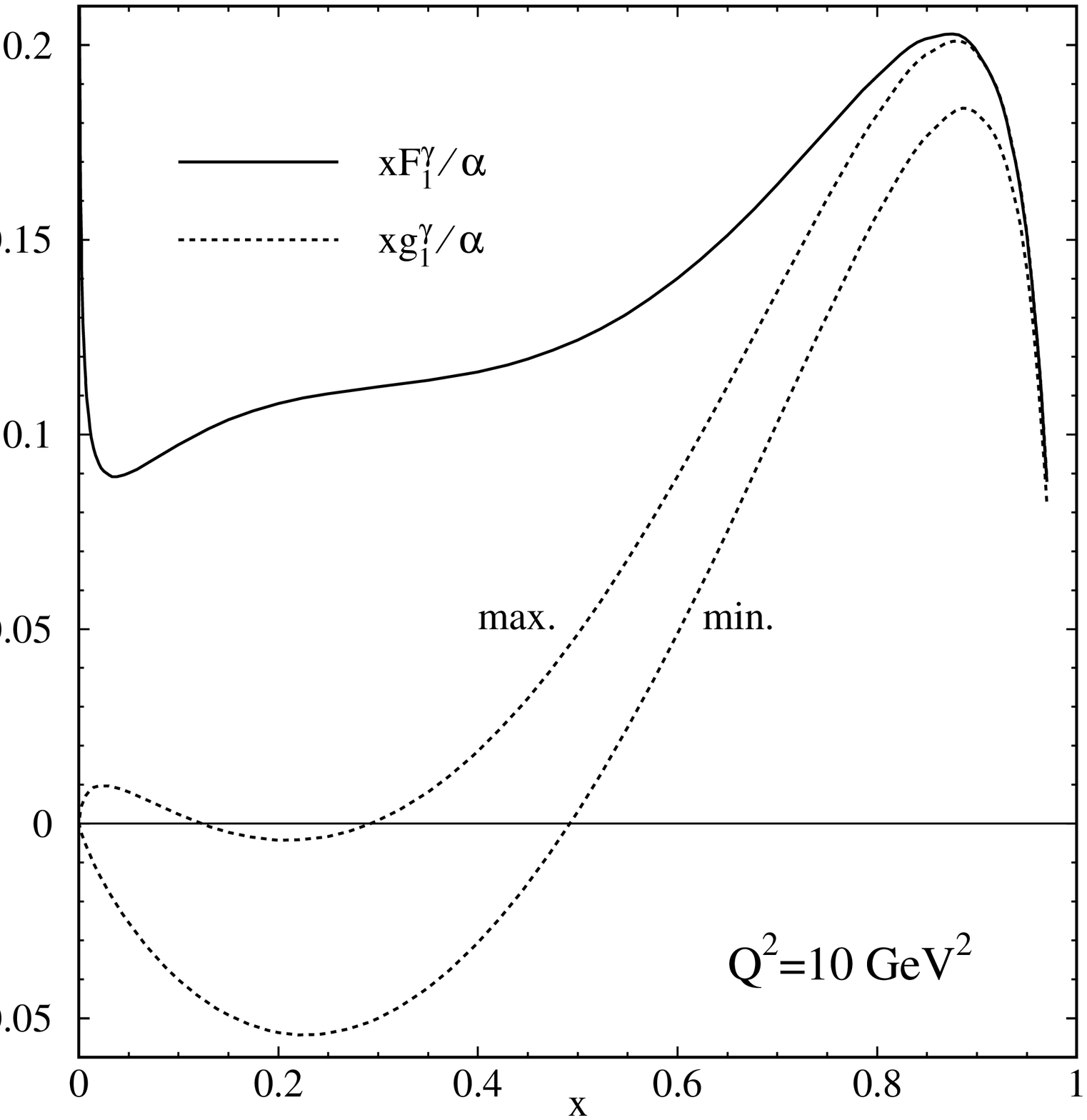,height=\textwidth}
\vspace*{1cm}
\\{\large\bf Fig. 1}
\end{center}
\end{figure}

\pagebreak

\begin{figure}
\begin{center}
\epsfig{file=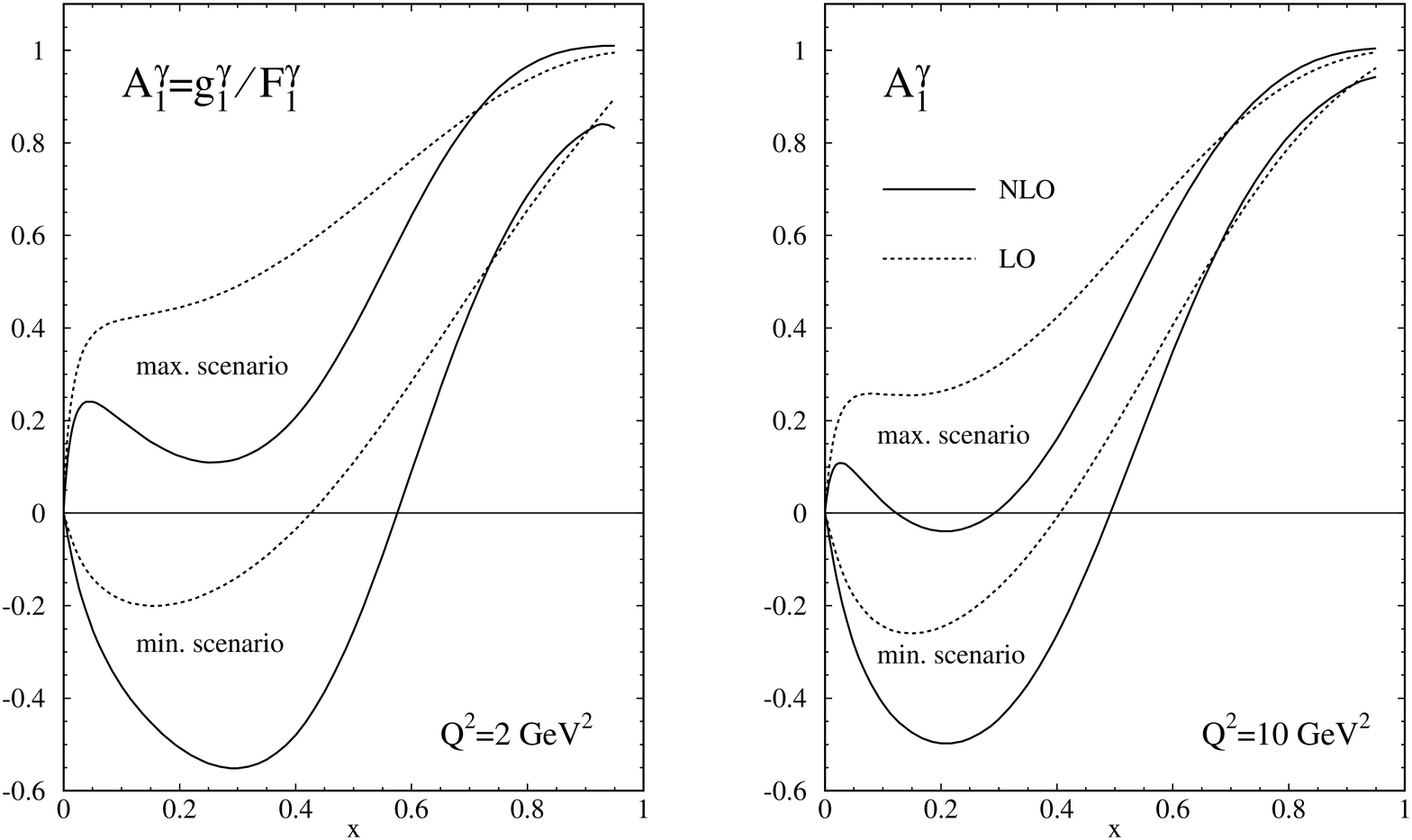,height=0.9\textwidth,angle=90}
\put(-5.,325.){\rotatebox{90}{\large\bf Fig. 2}}
\end{center}
\end{figure}

\pagebreak

\begin{figure}
\begin{center}
\epsfig{file=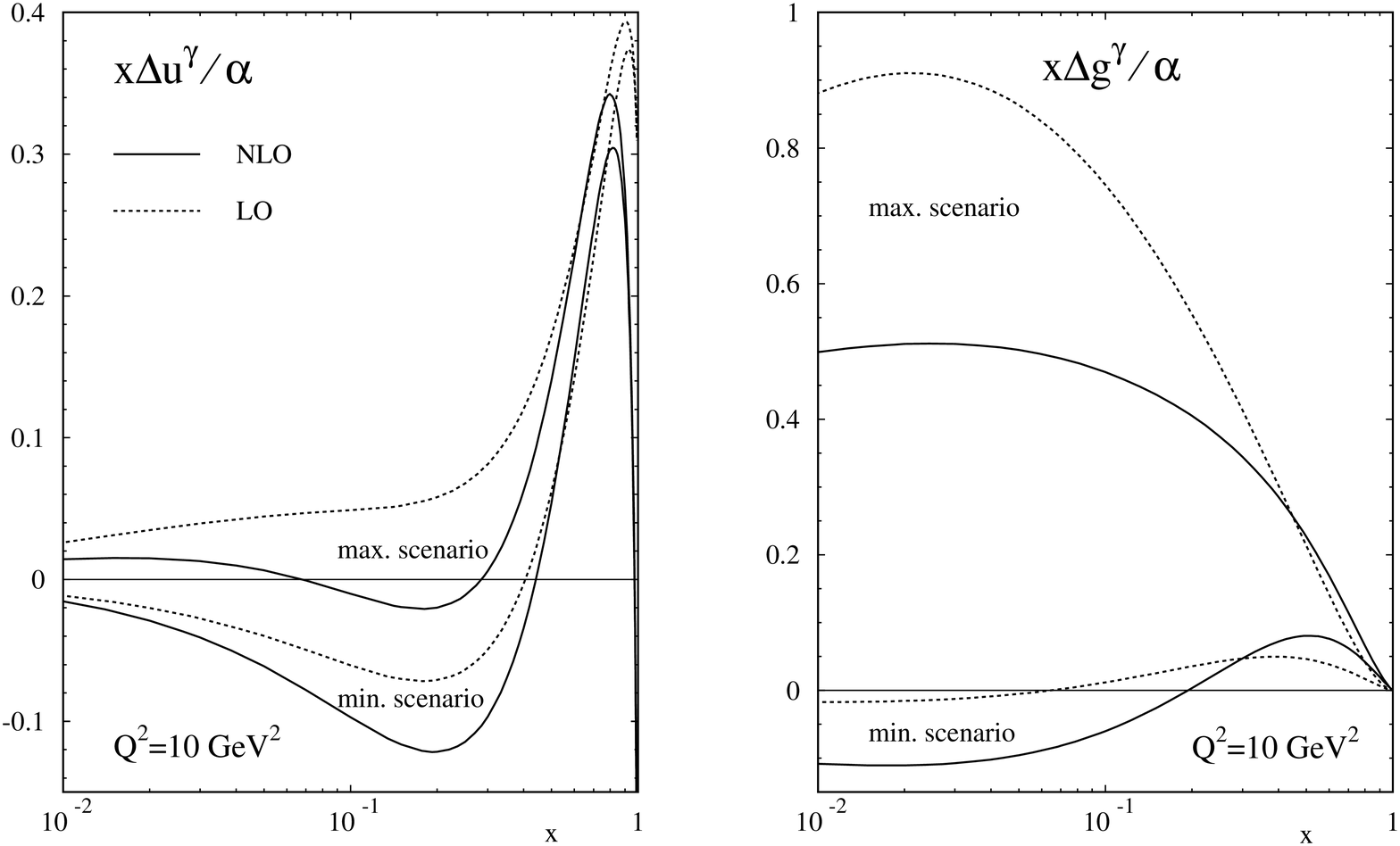,height=0.9\textwidth,angle=90}
\put(-5.,325.){\rotatebox{90}{\large\bf Fig. 3}}
\end{center}
\end{figure}

\pagebreak

\begin{figure}
\begin{center}
\epsfig{file=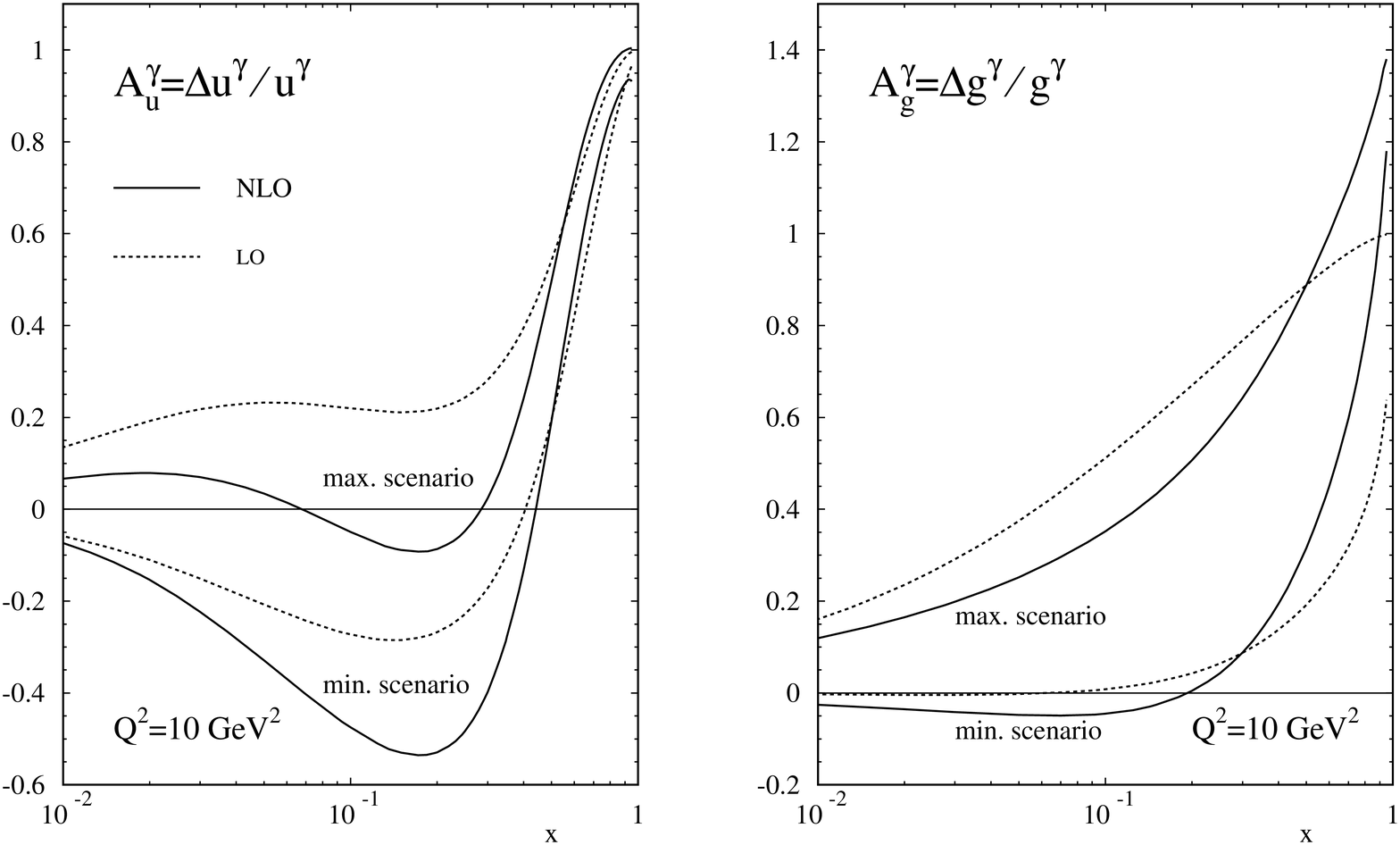,height=0.9\textwidth,angle=90}
\put(-5.,325.){\rotatebox{90}{\large\bf Fig. 4}}
\end{center}
\end{figure}

\end{document}